\documentclass[twocolumn,showpacs,preprintnumbers,amsmath,amssymb,floatfix]{revtex4}   
\usepackage{graphicx} 
\usepackage{dcolumn} 
\usepackage{bm} 
\begin{document} 
  
\title{Elastic scattering, inelastic excitation and 1$n$ pick up transfer cross sections for $^{10}$B + $^{120}$Sn at energies around the Coulomb barrier.}
      
\author{M. A. G. Alvarez\footnote{email: malvarez@us.es (corresponding author)}, M. Rodr\'{i}guez-Gallardo}
\affiliation{Departamento FAMN, Universidad de Sevilla, Apartado 1065, 41080 Sevilla, Spain}
\author{L. R. Gasques, L. C. Chamon, J. R. B. Oliveira, V. Scarduelli, A. S. 
Freitas, E. S. Rossi Jr.}    
\affiliation{Instituto de F\'{i}sica da Universidade de S\~ao Paulo, 05508-090, S\~ao Paulo, SP, Brazil} 
\author{V. A. B. Zagatto, J. Rangel, J. Lubian}
\affiliation{Instituto de F\'{i}sica da Universidade Federal Fluminense, 24210-346, Niter\'oi, Rio de Janeiro, Brazil}
\author{I. Padron}
\affiliation{Centro de Aplicaciones Tecnol\'ogicas y Desarrollo Nuclear, 502, Calle 30, La Habana, Cuba}

\date{\today}     
 
\begin{abstract} 
The $^{10}$B + $^{120}$Sn reaction has been systematically studied at laboratory
energies around the Coulomb barrier: $E_{\rm LAB}=$ 31.5, 33.5, 35.0 and 37.5 MeV. 
Cross sections for the elastic scattering and some reaction processes have been
measured: excitation to the 1$^+$ state of $^{10}$B; excitation to the 2$^+$ and 
3$^-$ states of $^{120}$Sn; and the 1 neutron pick up transfer 
$^{120}$Sn($^{10}$B,$^{11}$B)$^{119}$Sn. Coupled Reaction Channels (CRC) 
calculations have been performed in the context of the double-folding 
S\~ao Paulo Potential. The theoretical calculations result on a good overall 
description of the experimental angular distributions. The effect on the 
theoretical elastic scattering angular distributions of couplings to the 
inelastic and transfer states (through the CRC calculations) and to the 
continuum states (through Continuum-Discretized Coupled-Channels calculations) 
has been investigated. 
\end{abstract}      
      
      
\maketitle      
      
\section{Introduction}      
\label{introduct}      
      
\indent Exotic and weakly-bound nuclei originated in the primordial nucleosynthesis. 
In particular, the exotic $^{7}$Be and the stable weakly-bound $^{7}$Li were 
formed with very small yields. Nuclei heavier than $^{7}$Be and $^{7}$Li have 
been formed much later in the stellar nucleosynthesis, through light nuclei 
(H, He, Li, Be, B, \textit{etc}) reacting during stars evolution or explosion. 
Thus, the interstellar medium contains declining abundances of the light nuclei. 
On the other hand, the yields of light nuclei in the universe are also influenced by 
cosmic ray (mainly high-energy protons) spallation. This process is responsible 
for breaking the heavy elements up in the interstellar medium. In such scenario, 
any model to explain the current abundances of light nuclei must connect the 
stellar nucleosynthesis with the primordial one \cite{cha04}.

The study of exotic and stable weakly-bound nuclei is one of the forefronts 
of current research in nuclear physics \cite{tan13}. Recently developed facilities, 
worldwide, that produce radioactive ion beams, provide opportunities to probe 
new aspects of nuclear physics \cite{gal07,mot07,lep14} and astrophysics \cite{ber10}. The discovery of halo 
nuclei revealed that some exotic nuclei could have an extraordinary size \cite{tan85}. This 
discovery triggered many experimental and theoretical works to search for nuclei 
with unusual properties, such as an anomalous large radius or enhanced breakup 
cross sections. Nowadays, several nuclei, such as $^{6}$He, $^{11}$Li and 
$^{11}$Be, are well known to present a halo structure, where a core is 
surrounded by one or two weakly-bound nucleons, giving rise to a diffuse matter 
distribution that can produce an enhanced breakup cross section, even well 
below the Coulomb barrier \cite{maz06,esc07,san08,aco09,aco11,cub12,pie10,fer13,fer15}. In addition, reactions of exotic and stable 
weakly-bound nuclei at energies around the Coulomb barrier revealed the 
importance of the corresponding structure in the dynamics of the reaction 
processes, since it provides insight into degrees of freedom connected to slow 
processes.

Studying reactions involving weakly-bound stable nuclei is a crucial step 
towards a better understanding of the exotic ones. Comparing them is important 
for systematic studies with the purpose of determining how these nuclei react, 
aiming also to understand their respective abundances. Exotic and stable 
weakly-bound nuclei have two common fundamental characteristics: low breakup 
threshold and cluster structure. The breakup when interacting with another nucleus gives rise 
to a complex problem of three or more bodies. Breakup can occur by direct
excitation to the 
weakly-bound projectile into continuum states, or by populating continuum states of the  
target \cite{esc07,san08,aco09,aco11,raf10,luo11,kal16}. Close to or even below the Coulomb barrier, the Coulomb breakup even 
dominates some reactions of exotic nuclei with heavy targets \cite{fer15-prc}. In nuclear astrophysics, 
Coulomb breakup of weakly-bound projectiles has been used as an indirect method for 
determining cross sections of radioactive capture processes \cite{bot17,ver83,cha10}.

Weakly-bound stable nuclei can easily be produced and accelerated in conventional particle 
accelerators, where the reactions with several targets allow systematic studies with high statistics.
Within this context, the E-125 experimental campaign has been developed at the 
Open Laboratory of 
Nuclear Physics (LAFN, acronym in Portuguese) in the Institute of Physics of the University of 
S\~ao Paulo (IFUSP, acronym in Portuguese). The aim of the project is to study 
the scattering involving the light weakly-bound stable nuclei $^{6}$Li, 
$^{7}$Li, $^{9}$Be, $^{10}$B and $^{11}$B, on the same heavy target 
($^{120}$Sn), at energies around the Coulomb barriers, and compare the 
respective results among them as well as with others reactions involving exotic 
nuclei $^{6,8}$He, $^{8,9,11}$Li, $^{10,11}$Be. These reactions with the 
$^{120}$Sn target have produced many channels of inelastic excitation and 
transfer, which can be conveniently separated in the experimental energy 
spectra \cite{zag17,gas18}. 
The analyses of the different nuclear reactions processes, at several energies
and within the same theoretical approach, represent a powerful breakthrough. 

For $^{7}$Li + $^{120}$Sn, we have measured the elastic scattering, the
excitation to the 1/2$^-$  $^{7}$Li first excited state (E* = 478 keV), the
excitation to the 2$^+$ and 3$^-$ $^{120}$Sn states (E* = 1171 and 2400 keV)
and the one neutron stripping reaction, at energies close to the barrier 
($V_B({\rm LAB}) \approx 20.6$ MeV): $E_{\rm LAB}=$ 20, 22, 24 and 26 MeV \cite{zag17}. 
For this system, Coupled Reaction Channels (CRC) calculations have been performed 
in the context of the double-folding S\~ao Paulo Potential (SPP) \cite{cha02,alv03}. It turned out that 
the inclusion of the 1/2$^-$  $^{7}$Li first excited state as well as the projectile coupling 
to the continuum ($\alpha$ plus a tritium particle), play a fundamental role on the simultaneous 
description of the elastic, inelastic and transfer cross sections. Particularly, the simulation of the 
breakup effect, through the Trivial Local Equivalent Potential, suggested the importance 
of couplings to the continuum. On the contrary, the coupling to the one-neutron stripping channel 
does not significantly affect the theoretical elastic and inelastic scattering angular 
distributions.

Similar as $^{7}$Li, $^{10}$B also presents the first excited state with low excitation energy 
(1$^+$, 718 keV). Furthermore, it is a weakly-bound stable nucleus that may breakup into different mass partitions, 
being the most energetically favorable the $^{10}$B $\rightarrow$ $^{6}$Li + $^{4}$He (Q = -4.461 MeV). 
In such scenario, it could be important to include couplings to continuum states in order to describe simultaneously the 
elastic scattering and the different reaction channels. As part of the E-125 
experimental campaign, the $^{10}$B + $^{120}$Sn reaction has already been 
measured at 37.5 MeV and analysed within the CRC formalism \cite{gas18}. 
In the present work, we report on new experimental results obtained for the 
same reaction measured at $E_{\rm LAB} =$ 31.5, 33.5 and 35.0 MeV. Here, we 
present theoretical results obtained through CRC calculations and also those 
within the context of the Continuum-Discretized Coupled-Channels (CDCC).  

In the next section, the experimental setup presented in \cite{zag17,gas18} is
revisited. Then, we present the experimental data and respective 
theoretical analyses. Finally, we discuss our results and present the main 
conclusions.  
          
            
\section{Experimental Setup}      
\label{setup}      

\indent The development of new instrumentation has allowed more complex nuclear 
reaction measurements 
in the LAFN. The results of such experiments seek to answer some of the 
relevant questions regarding the understanding of cluster-like properties of 
light nuclei and how these properties influence reactions.

As previously mentioned, the experiment $^{10}$B + $^{120}$Sn is part of the E-125 campaign that has been 
carried out at LAFN. The experimental setup is based on SATURN (Silicon Array based on Telescopes of 
USP for Reactions and Nuclear applications). SATURN is being developed as a portable nuclear reaction 
spectrometer, based on silicon detectors telescopes, desktop-type electronic modules, and multi-channel 
acquisition systems. SATURN is installed in the 30B experimental beam line of the laboratory, which 
contains a scattering chamber connected to the 8MV pelletron tandem accelerator
\cite{zag17,gas18}. 

The reaction $^{10}$B + $^{120}$Sn was measured at the bombarding energies 
$E_{LAB}=$ 31.5, 33.5, 35.0 and 37.5 MeV. The 
SATURN detecting system was mounted with 9 surface barrier detectors in angular intervals of 5$^{\rm o}$, covering 40$^{\rm o}$ in 
each run. Normally, with 3 runs we cover approximately 120$^{\rm o}$, from 
40$^{\rm o}$ to 160$^{\rm o}$. The energy 
calibration of each detector was performed following the same procedure adopted
in \cite{gas18}.

As illustration, a typical spectrum taken at $E_{\rm LAB} = 35$ MeV and 
$\theta_{\rm LAB} = 125^{\rm o}$ is 
shown in Fig. 1. All the peaks have been identified and labeled using different 
colors. The elastic 
scattering peak of $^{10}$B incident on $^{120}$Sn is labeled in orange. The 
peak relative to the 
1$^+$ $^{10}$B inelastic excitation is given in red, while the excitation to 
the 2$^+$ and 3$^-$ 
$^{120}$Sn states are indicated by the green arrows. The peaks corresponding to different energy 
levels of the 1n pick up transfer are indicated by the blue arrows. For this reaction, different 
states of the compound $^{119}$Sn nucleus are separated and integrated in two groups. These groups 
can be identified in the spectrum of Fig. 1: the first group refers to the 
ground-state (g.s.), 23.8 and 89.5 keV excited states, and the second group 
refers to the 787.0, 920.5, 921.4, 1089 and 1354 keV excited states.

\begin{figure}[ht]  
\begin{center}  
\includegraphics[width=8.7cm]{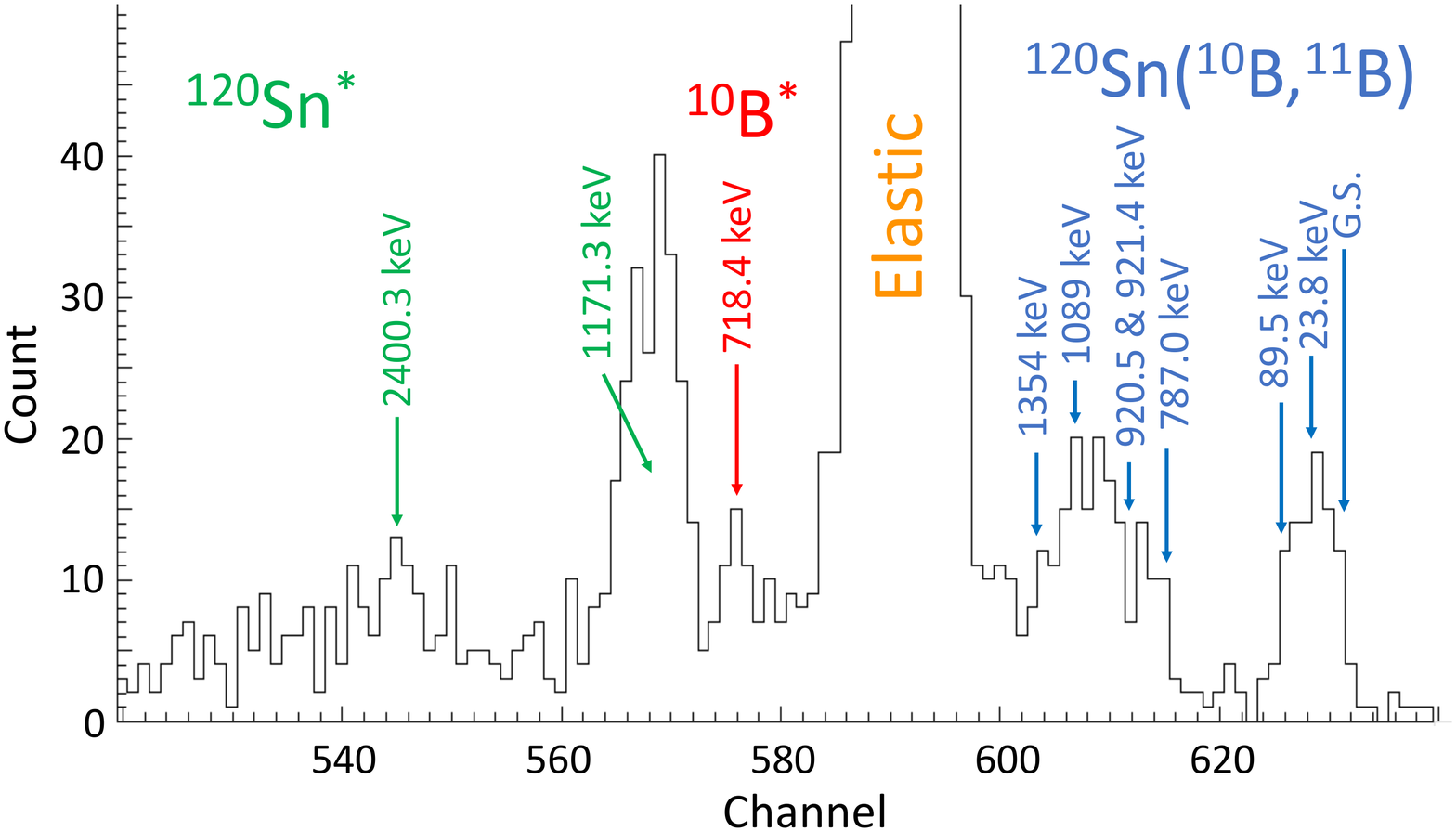}  
\caption{(Color online) Spectrum taken at $\theta_{\rm LAB}$ = 125$^\circ$ and $E_{\rm LAB} = 35$ MeV. 
The peaks corresponding to the 1$n$ pick up transfer (blue), elastic scattering of $^{10}$B on $^{120}$Sn (orange), 
inelastic excitation to the 1$^+$ $^{10}$B first excited state (red), and the
excitation to the 2$^+$ and 
3$^-$ $^{120}$Sn states (green), can be clearly identified in the figure.}  
\label{spectrum}  
\end{center}  
\end{figure} 


\section{Experimental Data and CRC Calculations}      
\label{theo}      

\indent In this section, we present an analysis of the elastic scattering and 
reaction channels. With this 
aim, we have performed CRC calculations as well as (elastic scattering) 
single-channel optical model (OM) calculations. The SPP is assumed for the real 
part of the optical potential. 
For the imaginary part, we have assumed the SPP multiplied by a fixed 
normalization factor, $N_I$ = 0.25, that has provided the best results in the 
previous CRC data analyses for $E_{\rm LAB} =$ 37.5 MeV \cite{gas18}. The FRESCO code 
was used to calculate the theoretical cross sections. As in \cite{gas18}, the 
collective vibrational mode was assumed to describe the quadrupole and octupole 
excited states of the $^{120}$Sn, whereas the $^{10}$B was treated as a rotor. 
All the parameter values assumed in the present CRC calculations (which are 
provided in tables I, II and III of Ref. \cite{gas18}) have been determined in the 
previous data analyses for $E_{\rm LAB} =$ 37.5 MeV. In this sense, no 
adjustable parameters are involved in the description of the data for the other 
energies. Thus, for $E_{\rm LAB}=$ 31.5, 33.5 and 35.0 MeV we, in fact, deal 
with theoretical predictions instead of data fits.

\subsection{Elastic Scattering}
\label{elas-scat}

Fig. 2 
presents experimental data and theoretical calculations for the elastic scattering 
angular distributions at the bombarding energies of 31.5, 33.5, 35.0 and 37.5 MeV.  
The red and blue lines in the figure refer to OM and CRC calculations, 
respectively. Since both kinds of calculations were performed with the same 
optical potential, the difference between the curves directly translates the 
effect of the couplings. 

The effect of the couplings is not much significant. Both calculations (CRC and 
OM) reproduce reasonably the data for all energies. Unlike other energies, the 
experimental data at 33.5 MeV present a Fresnel peak around 100$^\circ$, which 
is not reproduced by the theoretical calculations. On the other hand, the 
pronounced Fresnel peak that is present in the theoretical angular distributions 
at 37.5 MeV is absent from the data. Thus, the CRC calculations do not 
reproduce the data taken at 33.5 and 37.5 MeV around the Fresnel angular region. 
As $^{10}$B can be considered a cluster formed by $^4$He + $^6$Li, it is 
important to test the effect of the couplings to continuum states. This is the 
subject of section IV.

\begin{figure}[ht]  
\begin{center}  
\vspace{-0.3cm}
\includegraphics[width=8.5cm]{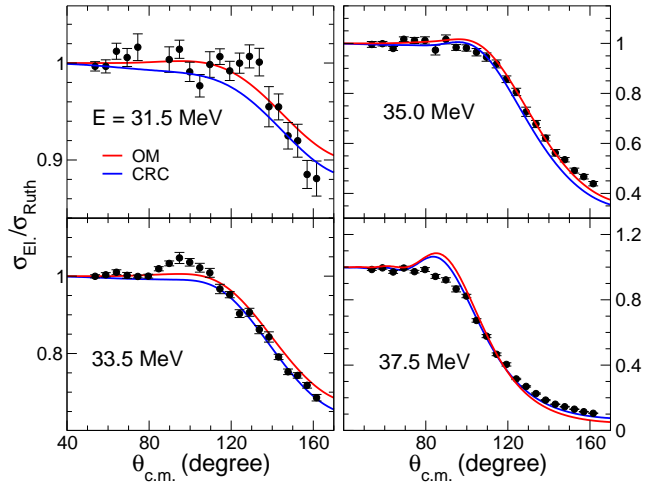}  
\caption{(Color online) Experimental and theoretical results for the elastic 
scattering angular 
distributions of $^{10}$B + $^{120}$Sn at different bombarding energies.}  
\label{elas}  
\end{center}  
\end{figure}

\subsection{Inelastic Excitation}
\label{inel-exc}

Fig. 3 
presents the cross sections for the inelastic excitation
of the 1$^+$ state of $^{10}$B (E$^*$ = 0.718 MeV) at 33.5 and 35.0 MeV, while 
Fig. 4 
shows the angular distributions for the excitation to the 2$^+$ $^{120}$Sn 
state (E$^*$ = 1.171 MeV) at 31.5, 33.5 and 35.0 MeV. The results of the CRC 
calculations are in good agreement with the data for all these cases.
Fig. 5 
presents the angular distributions for the inelastic excitation to the 
3$^-$ $^{120}$Sn state (E$^*$ = 2.400 MeV) at 33.5 and 35.0 MeV. Again, data 
and CRC theoretical calculations are compatible within the error bars. However,
as already discussed in \cite{gas18}, we point out that the data for inelastic 
excitation may have some contamination (in the experimental spectra) related to 
the 1n transfer process populating high excited states of $^{119}$Sn. 

\begin{figure}[ht]  
\begin{center}  
\vspace{-0.3cm}
\includegraphics[width=8.5cm]{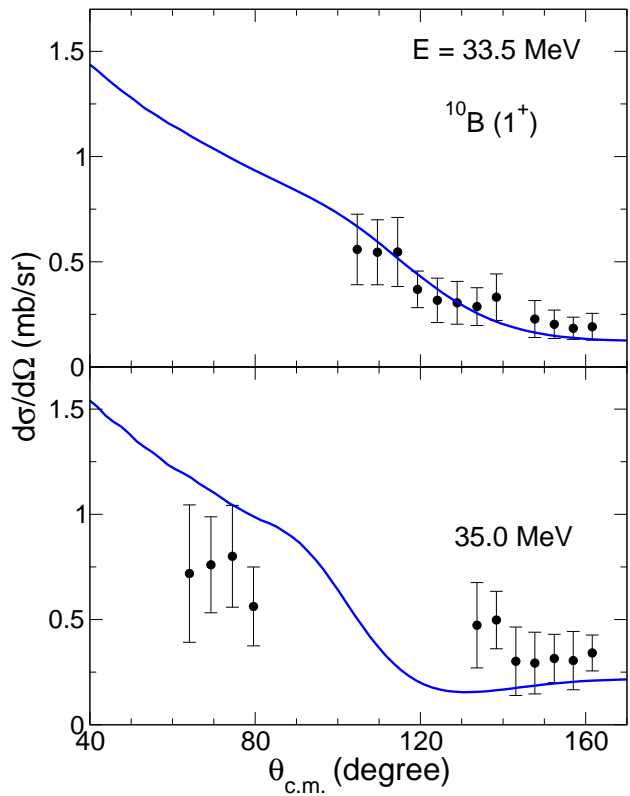}  
\caption{(Color online) Inelastic scattering angular distributions for the quadrupole 
excitation in $^{10}$B.}  
\label{inel1}  
\end{center}  
\end{figure}

\begin{figure}[ht]  
\begin{center}  
\vspace{-0.6cm}
\includegraphics[width=8.5cm]{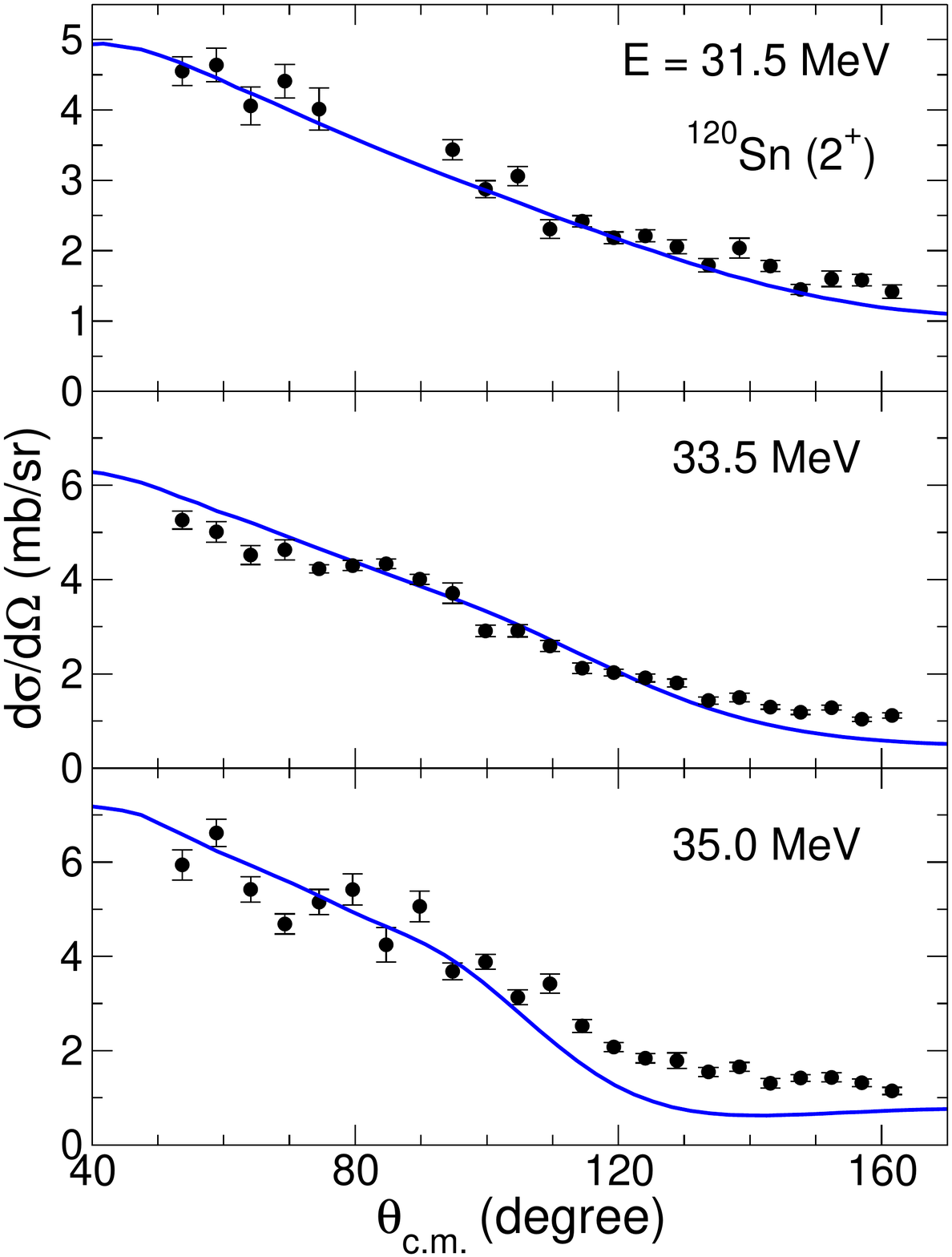}  
\caption{(Color online) Inelastic scattering angular distributions for the quadrupole 
excitation in $^{120}$Sn.} 
\label{inel2}  
\end{center}  
\end{figure}

\begin{figure}[ht]  
\begin{center}  
\vspace{-0.6cm}
\includegraphics[width=8.5cm]{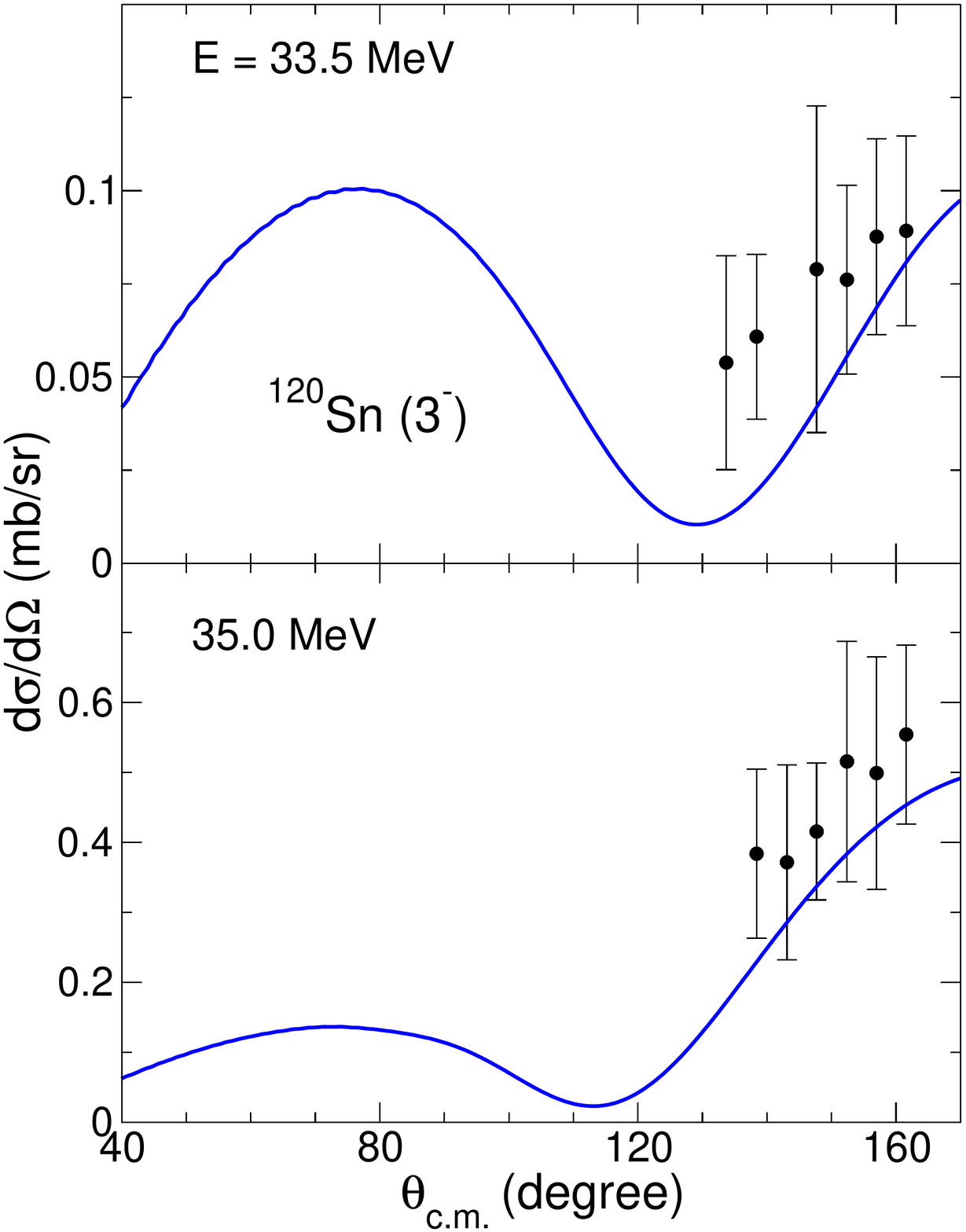}  
\caption{(Color online) Inelastic scattering angular distributions for the octupole 
excitation in $^{120}$Sn.} 
\label{inel3}  
\end{center}  
\end{figure}

\subsection{One-Neutron Transfer Cross Sections}
\label{1n-transfer}

\begin{figure}[ht]  
\begin{center}  
\vspace{-0.2cm}
\includegraphics[width=8.5cm]{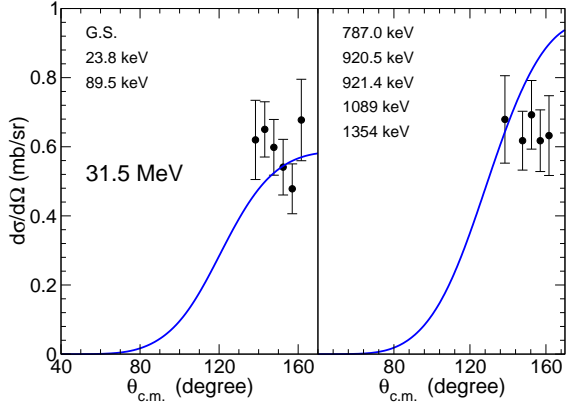}  
\caption{(Color online) 1$n$ pick up transfer angular distributions at 31.5 MeV.}  
\label{trans31}  
\end{center}  
\end{figure}

\begin{figure}[ht]  
\begin{center}  
\vspace{-0.2cm}
\includegraphics[width=8.5cm]{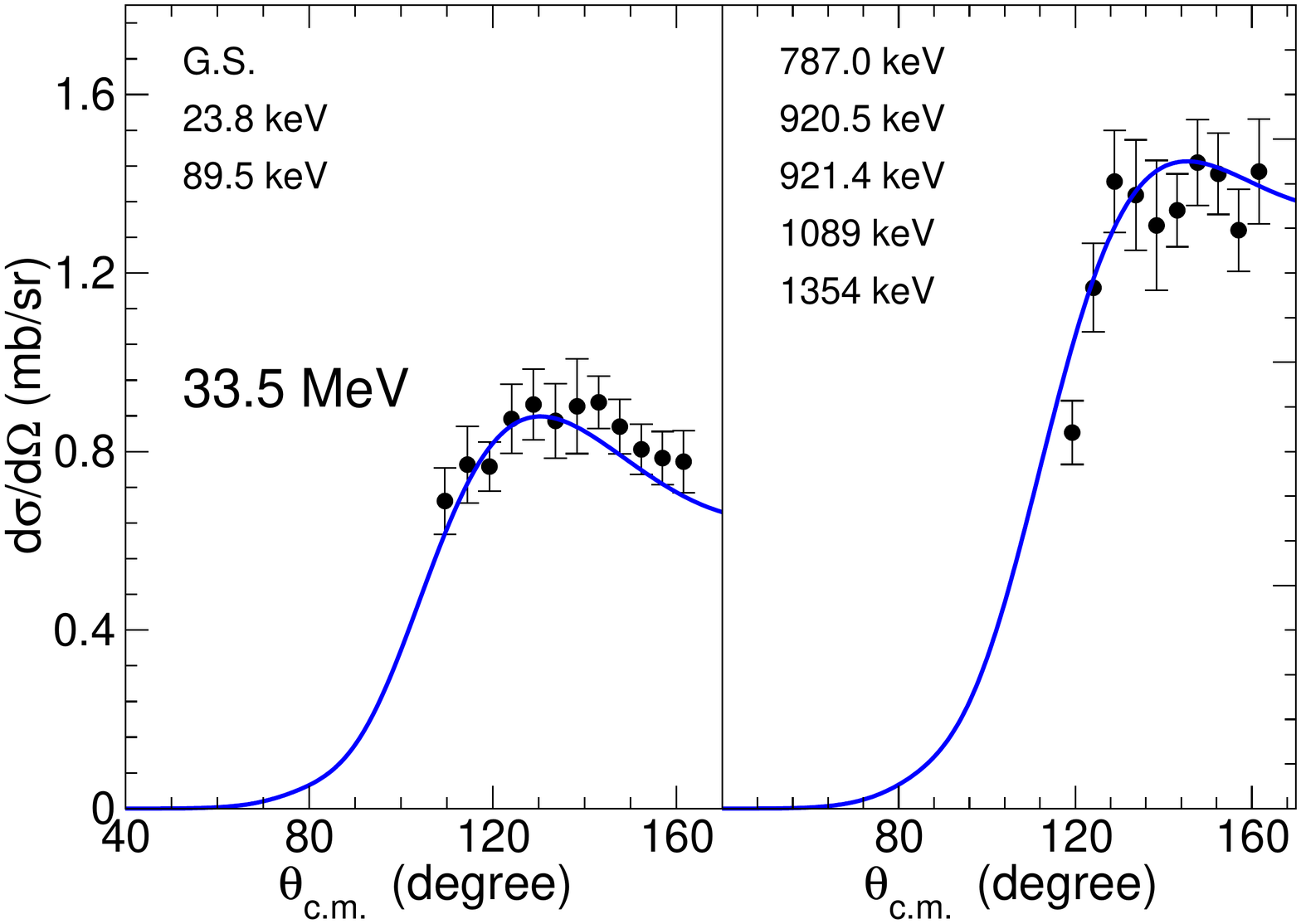}  
\caption{(Color online) Same as figure \ref{trans31} for 33.5 MeV.}  
\label{trans33}  
\end{center}  
\end{figure}

\begin{figure}[ht]  
\begin{center}  
\vspace{-0.2cm}
\includegraphics[width=8.5cm]{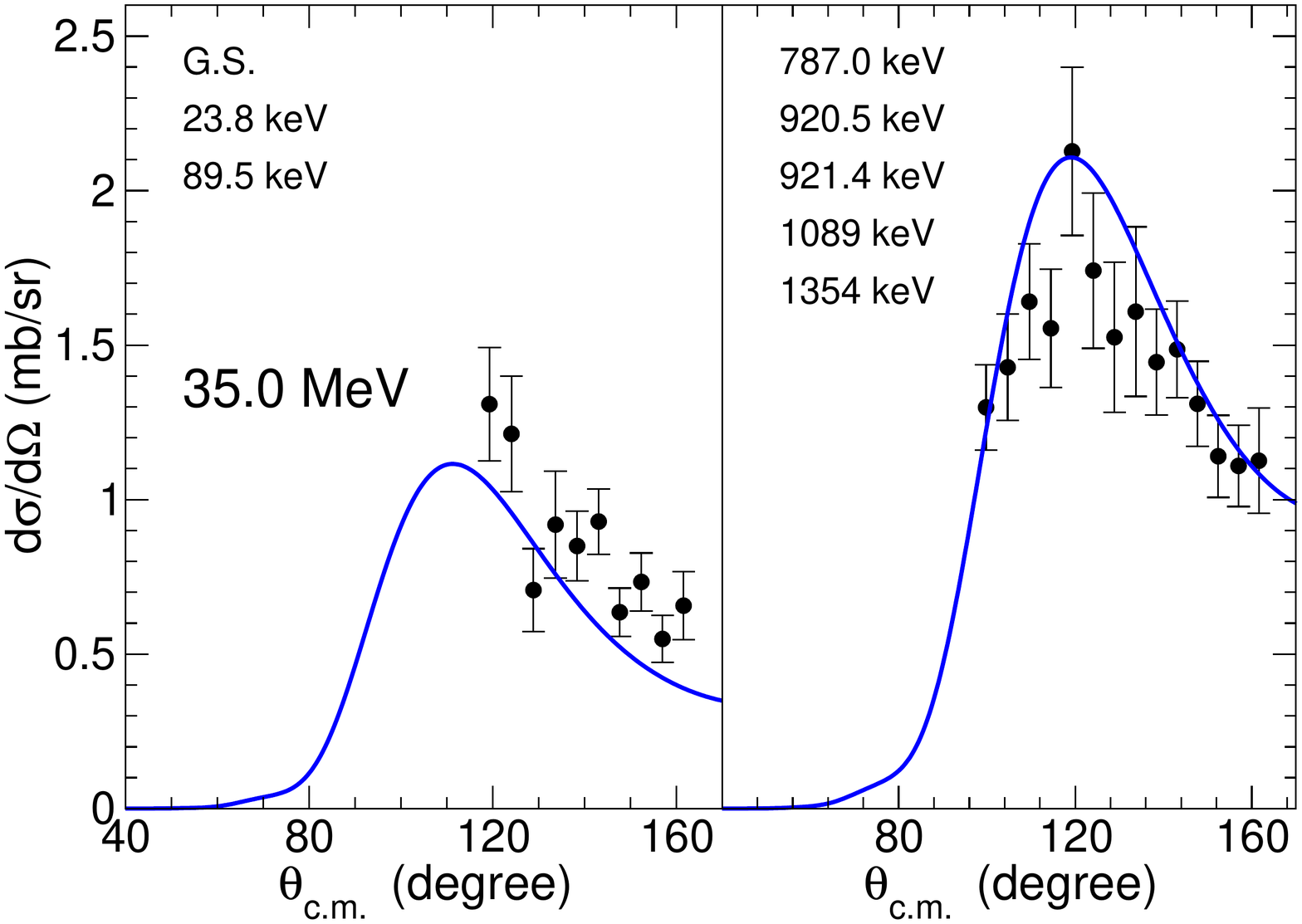}  
\caption{(Color online) Same as figure \ref{trans31} for 35.0 MeV.}  
\label{trans35}  
\end{center}  
\end{figure}

Experimental differential cross sections for 1$n$ pick up,
$^{120}$Sn($^{10}$B,$^{11}$B)$^{119}$Sn ($Q_{g.s.}$ = 2.350 MeV), have been 
obtained for different states of the residual $^{119}$Sn nucleus. As reported 
in \cite{gas18} for 37.5 MeV, the yields corresponding to different states of $^{119}$Sn 
could not be determined individually. In fact, two relatively broad peaks were 
observed in the spectra (see Fig. 1). The first group includes the g.s., 23.8 
and 89.5 keV states, whereas the second group includes the 787.0, 
920.5, 921.4, 1089, and 1354 keV states. 

Before presenting the analyses for the other energies, we comment the main 
results obtained (in Ref. \cite{gas18}) from the data analyses for 
$E_{\rm LAB} = 37.5$ MeV. The Woods-Saxon shape was assumed for the nuclear 
potentials of the neutron-core systems (n + $^{10}$B and n + $^{119}$Sn). The 
corresponding values for the radius, diffuseness, and depth for the g.s. of 
$^{11}$B and $^{120}$Sn are listed in the Table II of Ref. \cite{gas18}. The depth for 
each state is automatically adjusted by the FRESCO code in order to reproduce 
the corresponding biding energies. The inclusion of spin-orbit potentials for 
the n + $^{10}$B and n + $^{119}$Sn systems has a negligible effect on the 
calculated transfer angular distributions. The spectroscopic factor values 
assumed for states of the overlap $\langle$$^{120}$Sn $|$ $^{119}$Sn + 
n$\rangle$ are given in Table III of Ref. \cite{gas18}. 
For the overlap $\langle$$^{10}$B $|$ $^{9}$B + n$\rangle$, we have obtained 
several different sets of values for the spectroscopic factor and potential 
parameters that provide quite 
similar theoretical angular distributions for the neutron transfer process. 
This ambiguity is extensively discussed in Ref. \cite{gas18}.

In the present work, we have performed CRC calculations for all energies in  
the same conditions as reported in \cite{gas18}. The ambiguity observed in 37.5 MeV 
(commented in the previous paragraph) is also observed for the other energies. 
Figs. \ref{trans31}, \ref{trans33} and \ref{trans35} present experimental and 
theoretical (CRC) 1$n$ pick up transfer cross sections measured at 31.5, 33.5 
and 35.0 MeV. The cross sections correspond to the sum of the individual 
contributions relative to different excited states of $^{119}$Sn. Considering 
the lack of adjustable parameters, the agreement between data and theoretical 
results is remarkable for all energies. 

\section{Effect of the breakup channels on the elastic scattering}
\label{cdccc}

As described in the last section, the CRC calculations provide a quite
reasonable simultaneous description of the data relative to the elastic 
scattering, 
projectile and target excitations, and neutron transfer processes. The only 
significant disagreement between theory and experiment is observed for the
elastic scattering at the Fresnel angular region. Thus, we performed CDCC
calculations in order to verify whether that discrepancy would be related to the
couplings to the continuum states.

The binding energy of $^{10}$B is 4.461 MeV when it breaks into an alpha 
particle plus a $^6$Li nucleus. Despite having low breakup threshold of 1.473 
MeV, the sequential breakup of the $^6$Li fragment is not considered in our 
calculations since it is a second-order interaction. To account for the effect 
of the breakup channel on the elastic scattering angular distributions, we
have assumed the standard CDCC method \cite{sak86,raw74,raw75}, using the 
cluster model to describe the states of the projectile. This means that, in 
our CDCC calculations, we have not included the target excitations nor the 
transfer channels. In fact, our CRC calculations demonstrated that such
couplings do not have significant effect on the elastic scattering angular
distributions.

To perform CDCC calculations, one expands the total wave-function of the system, 
for the total angular momentum $J$ and projection $M$, in states of a basis as:
\begin{equation}
\Psi ^{JM}(\textbf{R}, \textbf{r}) = \sum_{\alpha} \, 
\frac{ f_{\alpha,J}(R)}{R} \,
\mathcal{Y}^{JM}_{\alpha}( \textbf{\^R},\textbf{r}) ,
\label{psi}
\end{equation}
where \textbf{r} represents the internal intrinsic coordinate of the projectile 
and \textbf{R} is the projectile-target relative coordinate. In eq. (\ref{psi}), 
$\mathcal{Y}^{JM}_{\alpha}( \textbf{\^R},\textbf{r})$ represents the tensor 
product of the angular part of the projectile-target relative wave-function 
with the intrinsic wave-function of the projectile. In this expansion, we 
use the binning method to generate the basis of square-integrable wave-functions 
of the projectile. They are obtained by taking the energy average of 
$^{6}$Li + $\alpha$ scattering states, within a given energy range (bin). They 
are then labeled by the mid point of the energy interval and by its angular 
momentum. We consider orbital angular momenta up to $l=3\,\hbar$. Using these 
states, the g.s. and the bound-states with energy lower than the binding energy
of the clusters, one builds an orthonormal basis to describe the continuum 
space of the projectile. The details of this procedure can be found in 
Refs. \cite{kam86,aus87}.

Inserting Eq.~(\ref{psi}) into the Schr\"odinger equation and carrying out some 
algebra, one obtains the following set of coupled equations:
\begin{eqnarray}
\nonumber
\Big[  H_\alpha\  - \,\left( E -\varepsilon_\alpha \right) \Big] \, 
f_{\alpha,J}(R) + \\
\sum_{\alpha' \ne \alpha} i^{L'- L} \, V_{\alpha \alpha'}(R) \, 
f_{\alpha', J}(R) = 0,
\label{CC}
\end{eqnarray}
where
\begin{equation}
\label{Halpha}
H_\alpha = -\frac{\hbar^2}{2\mu} \left[ \frac{d^2 }{dR^2}  - \frac{ L(L+1)}{R^2 }  \right] \,+\,V_{\alpha\alpha}(R)
\end{equation}
is the Hamiltonian in channel $\alpha$, and $\varepsilon_\alpha$ is the 
intrinsic energy of the projectile in this channel. In the present calculation, 
$\alpha=0$ stands for the elastic channel, where the projectile is in its 
g.s. ($\varepsilon_0=0, l_0=2, j_0=3$), and $\alpha \ne 0 $ corresponds to a 
projectile's state of an excited bound state or of a continuum bin state,
with energy $\varepsilon_{\alpha}$.

The matrix-elements of Eq.~(\ref{CC}) are given by
\begin{equation}
\label{matel}
 V_{\alpha \alpha '}(R)  = \langle \phi_{\alpha}({\bf r}) \left | V\left({\bf R},{\bf r}  \right)  \right | \phi_{\alpha'}(\bf{r}) \rangle,
\end{equation}
where $\phi_{\alpha}(\bf{r})$ stands for the wave-functions of both bound 
states and bins of the projectile.

The projectile-target interaction is given by the sum,
\begin{equation}
\label{eq4}
V \left( {\bf R},{\bf r}  \right)  =  V_{^4{\rm He} - ^{120}{\rm Sn}} \left(  
{\bf r}_{\rm v}   \right)  + V_{^6{\rm Li} - ^{120}{\rm Sn}} 
\left( {\bf r}_{\rm c} \right),
\end{equation}
where $V_{^4{\rm He} - ^{120}{\rm Sn}}$ and $V_{^6{\rm Li-^{120}\rm {Sn}}}$  are the 
optical potentials responsible for the elastic scattering of the valence 
particle ($^4$He) and of the core ($^ 6$Li) on the target ($^{120}$Sn). 
They are functions of the position vectors of the valence particle 
(${\bf r}_{\rm v}$) and the core (${\bf r}_{\rm c}$), respectively. These 
vectors are given in terms of the vector joining the centers of the 
collision partners (${\bf R}$) and the vector between the valence particle and 
the core (${\bf r}$), by the standard relations,
\begin{equation}
\label{transf}
{\bf r}_{\rm v} =  {\bf R} + \frac{A_{\rm c}}{A_p}\, {\bf r} \, \, {\rm and} 
\, \,  {\bf r}_{\rm c} =  {\bf R} - \frac{A_{\rm v}}{A_p} \, {\bf r},
\end{equation}
where $A_{\rm c},  A_{\rm v}$ and $A_p$ are the mass numbers of the core, the 
valence particle and the projectile, respectively.

The SPP was assumed for the real and imaginary parts of the 
$V_{^4{\rm He} - ^{120}{\rm Sn}}$ and $V_{^6{\rm Li-^{120}\rm {Sn}}}$
optical potentials. The strength coefficient of the imaginary part was set as 
$N_I = 0.78$. This strength coefficient has shown to be able to describe the 
elastic scattering of many systems involving tightly-bound nuclei, in wide mass 
and energy ranges \cite{alv03}. All intrinsic states of the projectile (bound 
and unbound) were determined by solving the Schr\"odinger equation for the 
$^6$Li + $\alpha$ system, assuming the SPP for the real part of the potential.

For the four energies studied in this work, the convergence of the CDCC method 
was checked in details to warrant that the results do not depend on the model 
space used. To solve the set of coupled equations (Eq.~(\ref{CC})), the 
matrix-elements $V_{\alpha\alpha'}(R)$ are expanded in multipoles up to 
$\lambda =3$, and their multipole components are evaluated by numerical 
integration over a mesh of radial distances (between the core and the valence 
particle) distributed between $r=0$ and $r_{\rm max}=80$ fm, with integration 
step size of 0.02 fm.  The coupled equations are then solved numerically 
considering projectile-target distances up to $R=1000$ fm and angular momenta 
up to 500 $\hbar$. The maximum bin energy for all the energies was 10 MeV. The 
width of the bins were of 2.0 MeV for the two lower bombarding energies, while 
it was of 1.5 MeV for the other two energies.
 
Data and theoretical results for the elastic scattering angular distributions, 
at the four bombarding energies, are presented in Fig. 9. The solid black lines 
in these figures correspond to the cross sections obtained with the CDCC 
calculations considering the couplings to the bound and continuum states, while 
the dashed red lines (labeled by no coupling) represent the CDCC results 
obtained turning off the couplings (remaining, therefore, only the g.s.). We 
point out that the no coupling calculations account for the cluster structure 
of the projectile in the g.s., i.e. the optical potential of this case 
corresponds to that given in expressions (\ref{matel}) and (\ref{eq4}), with 
$\alpha =0$.

\begin{figure}[ht]  
\begin{center}  
\vspace{-0.2cm}
\hspace{-0.9 cm}
\includegraphics[width=9.5cm]{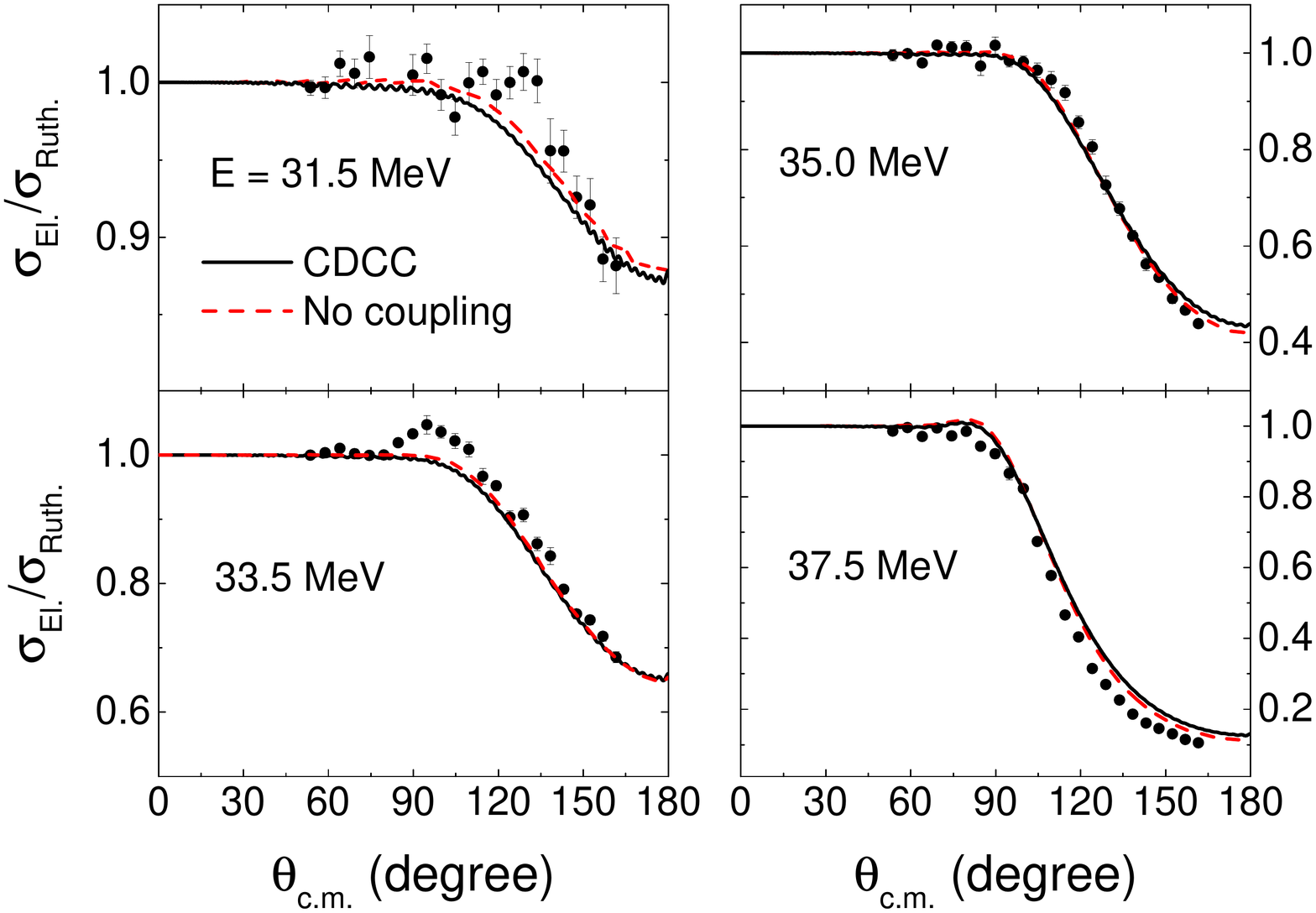}  
\caption{(Color online) Experimental and CDCC results for the elastic scattering angular
distributions. The no coupling case corresponds to the results of the CDCC 
calculations turning off the couplings.}  
\label{CDCCC}  
\end{center}  
\end{figure}

The no coupling theoretical cross sections presented in Fig. 9 are rather 
different from those of the OM in Fig. 2. In particular, the (theoretical) 
Fresnel peak at 37.5 MeV shown in Fig. 9 is much less pronounced than that in
Fig. 2. Both sets of theoretical cross sections are obtained from 
single-channel (g.s.) calculations but with different optical potentials. 
In fact, the results presented in Fig. 9 were obtained through Eqs. 
(\ref{matel}) and (\ref{eq4}), while those of Fig. 2 derive 
from the SPP. Of course, these two potentials should not be identical. 

The results presented in Fig. 9 indicate that the effect of the
couplings to the continuum on the elastic scattering cross sections is very
small. In particular, the discrepancy observed in Fig. 2 between theory and 
experiment at 33.5 MeV in the Fresnel angular region is still present 
in the results shown in Fig. 9 for the CDCC calculations. We mention that 
$^{10}$B can also be broken into p + $^9$Be and d + $^8$Be, with breakup 
thresholds larger than 6 MeV. As the 
binding energy is higher, one should expect that these two breakup modes should 
affect the elastic scattering distribution less than the breakup mode studied 
here.

Therefore, our theoretical calculations do not explain the behavior of the
elastic scattering angular distributions at the Fresnel region. On the other
hand, we have confidence that no significant systematic errors are present in 
the data set. In fact, the measurements for all energies, including 37.5 MeV, 
were performed in two consecutive weeks, using the same experimental setup and 
target. The setup consists of a dedicated nuclear reaction chamber,
vacuum, mechanics, electronics and data acquisition system, already used to
perform measurements for other systems, as, for instance, $^7$Li + $^{120}$Sn
\cite{zag17}. As reported in \cite{gas18}, we have used thin ($\approx 100$ 
$\mu$g/cm$^2$) isotopically enriched ($>$ 99\%) $^{120}$Sn targets, with a thin 
backing layer of $^{197}$Au for normalization purposes (at this energy range, 
the corresponding elastic scattering cross section is associated to the 
Rutherford one). With this, the cross sections can be accurately obtained, since 
they are related to the ratio between yields of $^{120}$Sn and $^{197}$Au. No 
trace of any kind of contaminants has been observed during the whole 
experimental campaign.

\section{Summary and Conclusion}      
\label{conclu}      

As part of the E-125 experimental campaign, that has been carried out at the 
LAFN, measurements for the $^{10}$B + $^{120}$Sn system have been performed at 
the bombarding energies of 31.5, 33.5, 35.0 and 37.5 MeV. Besides the elastic 
scattering channel, the excitation to the 1$^+$ state of $^{10}$B, the 
excitation to the 2$^+$ and 3$^-$ states of $^{120}$Sn, and the 1 neutron pick 
up transfer reaction $^{120}$Sn($^{10}$B,$^{11}$B)$^{119}$Sn have been observed.
The results corresponding to the measurement taken at 37.5 MeV were previously 
published in \cite{gas18}. A simultaneous analysis of the cross sections of 
these different channels has been performed within the CRC and OM formalisms in 
the context of the double-folding SPP interaction. The imaginary part of the 
optical potential that provides the best data fit for $E_{\rm LAB}= 37.5$ MeV 
was obtained in \cite{gas18}, by multiplying the SPP by the normalization factor of 
N$_I = 0.25$. This value is significantly smaller than that ($N_I = 0.78$)
obtained for many systems involving tightly-bound nuclei \cite{alv03}.

It is worth to mention that, for all energies, the CRC calculations were 
performed using the same set of parameter values for the nuclear potential, 
deformation lengths and spectroscopic factors. These values were determined in \cite{gas18}, 
through the analyses of the data for $E_{\rm LAB}= 37.5$ MeV. In this
sense, the data analyses for the other energies were performed here without
adjustable parameters.

The CRC calculations provide a good overall description of the complete
data set, except in the Fresnel angular region for the elastic scattering at  
33.5 and 37.5 MeV. In fact, the data at 33.5 MeV present a pronounced Fresnel 
peak which is not reproduced by the theoretical calculations. On the other hand, 
according to our theoretical approaches, a Fresnel peak is expected at 37.5 MeV 
around 100$^\circ$, but without correspondence in the experimental data. A
similar result has been recently reported in \cite{ara18} for the elastic scattering of 
$^9$Be on $^{120}$Sn. In this case, the Fresnel peak is prominent in the data 
at $E_{\rm LAB}=29.5$ MeV, damped at $E_{\rm LAB}=31$ MeV, and again appears in
higher energies (42 and 50 MeV). This behavior is not foreseen by the CDCC 
theoretical calculations reported in that paper.

In order to investigate the effect of the couplings to the continuum on the 
elastic scattering cross sections, we have also performed CDCC calculations 
considering the breakup of $^{10}$B into $^4$He plus $^6$Li. However, according 
to our theoretical results, this effect is quite small and, therefore, the 
discrepancy commented above between theory and experiment in the Fresnel region 
is still not understood. The discrepancy at the specific energy of
33.5 MeV is not related to any kind of contamination or normalization issue,
since no similar behavior is found in the data sets for the other energies. It 
would be of value to measure elastic scattering of $^{10}$B on $^{120}$Sn in 
higher energies, with the purpose of comparison with the behavior commented 
above for $^9$Be + $^{120}$Sn.

\begin{acknowledgments} 
This work has been partially supported by Funda\c{c}\~{a}o de Amparo \`{a}
Pesquisa do Estado de S\~{a}o Paulo (FAPESP), Conselho Nacional de
Desenvolvimento Cient\'{i}fico e Tecnol\'{o}gico (CNPq), 
Coordena\c{c}\~{a}o de Aperfei\c{c}oamento de Pessoal de N\'{i}vel Superior (CAPES), and 
Funda\c{c}\~{a}o de Amparo \`{a} Pesquisa do Estado do Rio de Janeiro (FAPERJ), 
Brazil. This work is a part of the project INCT-FNA Proc. $\rm N^o$ 464898/2014-5. 
This work has also been partially supported by the Spanish Ministerio de 
Econom\'{i}a y Competitividad, and the European Regional Development Fund (FEDER) 
under Project No. FIS2014-51941-P, by Junta de Andalucía under Group No. 
FQM-160, and by the European Union's Horizon 2020 research and innovation 
program under Grant Agreement No. 654002. M.A.G. Alvarez would like to thank 
the VI Plan Propio de Investigaci\'{o}n y Transferencia - Universidad de 
Sevilla (2017 and 2018).
\end{acknowledgments}      
  

\end{document}